\documentclass[a4,10pt,onecolumn]{article}
\usepackage{graphicx} 
\usepackage[margin=0.8in]{geometry} 
\usepackage{amsmath,amsthm,amssymb}
\usepackage{float}
\usepackage{xcolor}
\usepackage[utf8]{inputenc}
\usepackage{graphicx}
\usepackage[backend=bibtex,style=nature]{biblatex}
\usepackage{lineno}
\usepackage{authblk}

\def\_#1{{\bf #1\mit}}

\usepackage[font=footnotesize]{caption}

\title{Strong Mode Coupling via Quasi-Bound States in the Continuum in Bianisotropic Metasurfaces}

\date{}
\author[1]{L.M. Máñez-Espina$^*$}
\author[2]{B. Amrahi}
\author[2]{V.S. Asadchy}
\author[1]{A. Díaz-Rubio}
\affil[1]{Nanophotonics Technology Center, Universitat Politècnica de València, València 46022,  Spain.}
\affil[2]{Department of Electronics and Nanoengineering, Aalto University, Espoo, 02150, Finland.}

\begin{document}

\maketitle
\begin{abstract}
Electromagnetic mode coupling plays a key role in many resonant effects in nanophotonics. This coupling is also responsible for the appearance of bianisotropy, where electric and magnetic responses become interconnected through the interaction of their respective modes. In this work, we develop a simple and general temporal coupled-mode theory model to describe off-diagonal chiral bianisotropy. Using quasi-bound states in the continuum (q-BICs), we demonstrate how to control the hybridization of modes with opposite symmetries, resulting in Rabi-like splitting between the hybrid states in the regime of strong electromagnetic mode coupling. Beyond revealing the physical origin of the hybrid modes, our model predicts and explains the emergence of dual-band asymmetric reflection and absorption, and how to achieve maximum directional absorption difference. The theoretical predictions are verified by full-wave simulations, showing very good agreement with theory.  Furthermore, very strong reciprocal bianisotropy is demonstrated with the use of q-BICs in a deeply subwavelength metasurface in the optical frequency range. Our results provide a clear physical picture of the interaction process between modes, offering a compact theoretical framework for understanding and designing bianisotropic dielectric metasurfaces not only in the traditional regime but also in the strong coupling regime.
\end{abstract}

\section{\label{sec:Introduction} Introduction}

Resonant modes in nanophotonic structures are a central topic in modern optics because they enable light–matter interactions that can give rise to a wide range of exotic effects~\cite{Limonov2017,Babicheva2024,Absorption_Scattering_Small_Particles_book}. When the electromagnetic field remains confined for long times inside a resonator, even weak material responses or small geometrical asymmetries can produce large observable phenomena. This confinement is commonly quantified by the quality factor (Q-factor) of the resonance, which measures how long the energy can be stored before it is radiated or dissipated. Therefore, achieving high-Q resonances is essential for enhancing the interaction between light and structured matter~\cite{Review_USC,Guan2022,Baranov2018}.

A major breakthrough in the quest to obtain and control high-Q resonances in photonics was the introduction of bound states in the continuum (BICs)~\cite{Marinica2008}. BICs are non-leaky resonant modes that exist within the continuum of radiative states. These modes cannot be directly excited by external plane waves and, ideally, do not radiate energy into the surrounding medium. Their practical counterparts, known as quasi-bound states in the continuum (q-BICs), arise when a small perturbation allows these otherwise confined modes to couple to external excitation. These modes possess finite lifetimes, and their most remarkable feature is that their Q-factor can be tuned by controlling the strength of the perturbation~\cite{Koshelev2018}, providing a flexible mechanism for tailoring light confinement and radiative coupling~\cite{Kang2023}.

BICs can be classified according to their formation mechanism~\cite{Hsu2016,Zhen2014,Friedrich1985}. Among all types, the most relevant and widely studied in nanophotonics are the symmetry-protected BICs, particularly in dielectric metasurfaces~\cite{Huang2024}. In these cases, the symmetry of certain modes is incompatible with that of the incident field, preventing any radiative coupling. Breaking the protecting symmetry enables coupling with plane-wave excitation, transforming a BIC into a q-BIC. The resulting Q-factor scales inversely with the square of the perturbation strength~\cite{Koshelev2018,Overvig2020}, $Q=B/\delta^2$, where $B$ is a constant and $\delta$ is the characteristic size of the perturbation. Typically, in-plane symmetry breaking is employed to achieve this controlled coupling between normally incident light and q-BIC modes.

However, an even richer physical picture emerges when different resonant modes interact with each other~\cite{Liu2017}. The coupling between modes of distinct symmetry, polarization, or physical origin can profoundly modify the optical response of the system~\cite{Gladyshev2020, Poleva2023, Lepeshov2018}. Such mode–mode interactions can lead to hybridized states, energy splitting, and interference phenomena that cannot be understood from single-mode analysis~\cite{Zhang2025}. In the context of metasurfaces, most of the studies about coupling investigate the coupling between an excitonic mode of the material and a resonant electromagnetic mode supported by the metasurface~\cite{Dovzhenko2018,vandeGroep2013,Xie2021,Qin2023,Weber2023}. 
Recently, diagonal chirality, a reciprocal spatial effect that can be modeled as the electromagnetic coupling of the magnetic and electric dipoles that share directions in space, has been studied in conjunction with q-BICs~\cite{Kumar2025,Deng2025}, showing that maximum chirality can be achieved by coupling resonant modes.

Another interesting phenomenon arising from mode interaction is the electromagnetic coupling between \textit{orthogonal} electric and magnetic dipole moments. This effect is commonly referred to as omega coupling (off-diagonal chirality) due to the $\Omega$-shape of the metallic resonator, which intuitively explains this phenomenon~\cite{Tretyakov1993,Simovski1997, LibroAzul,Asadchy2018}. In optics, the same effect is usually realized by breaking out-of-plane symmetry with the use of a substrate or a two-material system resonator~\cite{Albooyeh2015,Fan2025}. This effect has not been, to the best of our knowledge, studied in conjunction with BICs.
 
In this work, we propose a theoretical framework based on temporal coupled-mode theory (TCMT) to describe and design optical metasurfaces that support maximum directional absorption via strong omega bianisotropy. This TCMT gives us insight into the possibilities of a dielectric structure that supports two opposite symmetry resonances and that has broken out-of-plane symmetry, which allows electromagnetic coupling. Interestingly, the TCMT can show the Rabi-splitting that strongly coupled resonances follow.  Furthermore, if losses are added to the system, new venues for optical responses open. We report a dual-band-like asymmetric absorption, similar to what can be implemented with metallic structures~\cite{Yazdi2015,Radi2013}, but with opposite responses between counter-propagating excitations, directional absorption difference, and out-of-band transparency. Moreover, the conjunction of q-BICs and broken out-of-plane symmetry allows for very strong and tunable off-diagonal bianisotropy in the optical regime.

\section{\label{sec:Bianisotropy and Hybridization}Omega bianisotropy and mode hybridization}

\begin{figure*}[t]
    \centering
    \includegraphics[width=1\linewidth]{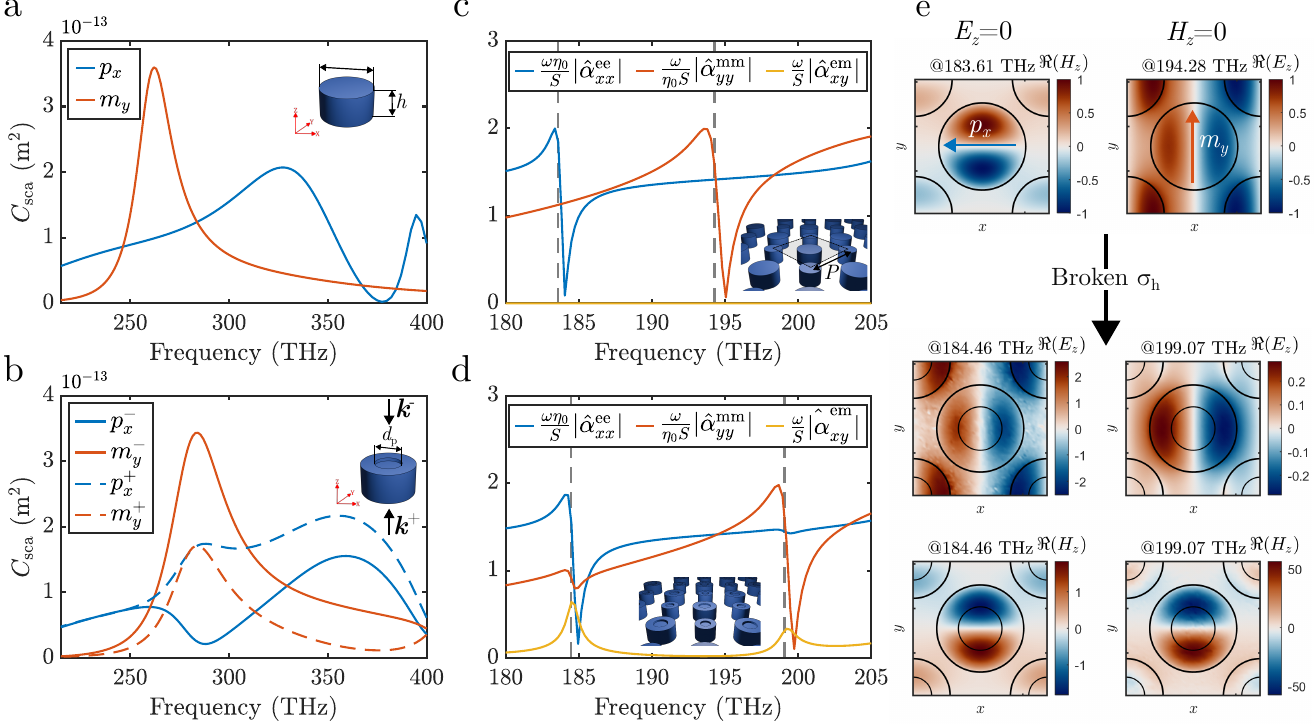}
    \caption{\textbf{Scattering properties of off-diagonal chiral bianisotropic particles and metasurfaces.} (a) Scattering cross-section of a single cylindrical nanoparticle. (b) Scattering cross-section of a broken mirror symmetric particle. Different multipole contributions for different propagation directions. Geometrical parameters: $D=300$~nm, $h=250$~nm, and $n_{\rm d}=3.5$. (c) Absolute value of the normalized collective polarizabilities calculated for the symmetric metasurface. In the frequency range shown, the structure holds two q-BIC eigenmodes: a magnetic dipole and an electric dipole resonance. (d) A metasurface with broken out-of-plane symmetry, holding the new hybridized resonant modes. Omega type polarizability component, $|\hat{\alpha}_{xy}^{\rm em}|$, is different from zero. Nanodisks have diameters defined as $D_{\rm a}=D+\Delta/2$ and $D_{\rm b}=D-\Delta/2$. Geometrical parameters: $P=1000$ nm $D=600$~nm, $h=250$~nm, $\Delta=100$~nm, $n_{\rm d}=3.5$. The perturbation, i.e., the perforations on the top part of the nanodisks have dimensions $h_{\rm p}=20$~nm, $\sigma_{\rm p}=0.5$.}
    \label{fig:Figure 1}
\end{figure*}

Under plane-wave illumination, the scattering cross-section of a dielectric nanoparticle can be computed using the multipole expansion of scattered fields~\cite{alaee2018electromagnetic}. 
For a single cylindrical nanoparticle placed in air with diameter $D$, height $h$, and refractive index $n_{\rm d}$, the dipolar contributions to the scattering cross-section are shown in Fig.~\ref{fig:Figure 1}(a). The particle is illuminated by a plane wave propagating along the $z$-axis ($\mathbf{k}=k_z\mathbf{u_z}$, where $\mathbf{u_z}$ is the unit vector pointing in the positive $z$-axis direction), with the electric field polarized along the $x$-direction.
Under these conditions, a magnetic dipole moment is created in the $y$-direction (represented as $m_y$) and an electric dipole moment along the $x$-direction (represented as $p_x$).
The scattering cross-section is the same for $k_z>0$ and $k_z<0$, as the cylinder has mirror symmetry, $\sigma_{\rm h}$, with respect to the $xy$-plane. Furthermore, the uniaxial symmetry of the cylinder ensures the same scattering cross-section for all polarizations. In this case, bianisotropy is forbidden by symmetry.

However, in the case where $\sigma_{\rm h}$ is broken by perforating the top part of the cylinder with depth $h_{\rm p}$ and diameter $d_{\rm p}$, as shown in the inset of Fig.~\ref{fig:Figure 1}(b), bianisotropy is enabled. The scattering cross-sections for counterpropagating plane waves are shown in Fig.~\ref{fig:Figure 1}(b). It can clearly be seen how, at previously resonant frequencies for the electric and magnetic dipole moments, both contributions interact. To further illustrate the discussion, we present the model for the electric and magnetic dipole moments in terms of the incident fields and the individual polarizabilities of the nanoparticle as~\cite{Alaee2015}
\begin{align}\label{eq:Individual Polarizabilities1}
\frac{p_x^{\pm}}{\varepsilon_0} &= \alpha^{\rm ee}_{xx} E^x_{\text{i}} \,\pm\, \alpha^{\rm em}_{xy} \eta_0 H^y_{\text{i}}, \\ 
\eta_0 m_y^{\pm} &= \alpha^{\rm me}_{yx} E^x_{\text{i}} \,\pm\, \alpha^{\rm mm}_{yy} \eta_0 H^y_{\text{i}}.
\label{eq:Individual Polarizabilities2}\end{align}
Where $\alpha^{\rm ee}$ is the electric polarizability, $\alpha^{\rm mm}$ is the magnetic polarizability, $\alpha^{\rm em/me}$ is the electromagnetic and magnetoelectric polarizabilies, all calculated for an individual particle, $\eta_0$ is the impedance of free space, $\varepsilon_0$ is the permittivity of free space, and $E_{\rm i}$ and $H_{\rm i}$ are the electric and magnetic incident fields. 
As already mentioned, broken $\sigma_{\rm h}$  enables bianisotropy. Specifically, the cross components of the $\alpha^{\rm em/me}$  polarizabilities can be different from zero. Notice that, as no external bias, nonlinearity, or time modulation is occurring, the structure is symmetrical with respect to time reversal, and reciprocity is preserved. Reciprocity ensures an important symmetry constraint $\alpha^{\rm em}=-(\alpha^{\rm me})^T$, therefore $\alpha_{yx}^{\rm me}=-\alpha_{xy}^{\rm em}$.

The expressions in Eqs.~\eqref{eq:Individual Polarizabilities1}–\eqref{eq:Individual Polarizabilities2} describe the individual polarizabilities of a single particle in isolation. However, in a periodic array, the local field acting on each particle differs from the incident field due to the mutual interactions between all other scatterers. These interactions give rise to collective polarizabilities, which account for the multiple-scattering and lattice effects within the metasurface.
The collective polarizability tensor, $\hat{\alpha}$, can be understood as an effective response function that relates the averaged induced dipole moments to the macroscopic incident fields on the array incorporating both the intrinsic response of the individual nanoparticle and the electromagnetic coupling mediated by the lattice~\cite{LibroAzul}. The scattering parameter expressions in terms of the collective polarizabilities are~\cite{albooyeh2023classification}

\begin{align}
t_{xx}^{\pm} &= 1 - \frac{j\omega}{2S} 
\left( \eta_0  \hat{\alpha}_{xx}^{\mathrm{ee}} + \frac{\hat{\alpha}_{yy}^{\mathrm{mm}}}{\eta_0} \right), \\[6pt]
r_{xx}^{\pm} &= -\frac{j\omega}{2S} 
\left( \eta_0 \hat{\alpha}_{xx}^{\mathrm{ee}} - \frac{\hat{\alpha}_{yy}^{\mathrm{mm}}}{\eta_0} \,\pm\, 2 \hat{\alpha}_{xy}^{\mathrm{em}} \right),
\end{align}
where $S$ represents the area of a single unit cell.
Fig.~\ref{fig:Figure 1}(c) shows the normalized collective polarizabilities for a checkerboard-like squared array of cylinders with different diameters between nearest neighbours~\cite{Geometry-1,Hu2020,Mez-Espina2024,BIASFREE}. Diameters are defined as $D_{\rm a}=D+\Delta/2$ and $D_{\rm b}=D-\Delta/2$, where $\Delta$ denotes the geometrical perturbation parameter. The array is characterized by a height $h$ and period $P$. Indeed, the difference in diameter between nearest neighbours is the perturbation that brings the BIC to a q-BIC. This specific perturbation makes the lattice from monoatomic to diatomic, bringing the dark mode in the $X$ point of the unperturbed lattice to the $\Gamma$ point of the perturbed lattice. This manipulation is usually called First Brillouin zone folding~\cite{Overvig2020}. The structure is illuminated at normal incidence by an $x$ polarized plane wave. It can clearly be seen that at approximately 194~THz the magnetic polarizability undergoes a rapid change due to a magnetic dipole-like resonance, and that the same takes place at approximately 184~THz for the electric polarizability due to an electric dipole-like resonance. The field profiles and resonant frequencies of such resonant modes are shown in Fig.~\ref{fig:Figure 1}(e). Due to $\sigma_{\rm h}$ symmetry, the eigenmodes must be either symmetric or antisymmetric with respect to $z$ axis. This will lead to a clear classification between modes that have either $H_z=0$ or $E_z=0$ in the $xy$-plane of symmetry. 
The first case will be that of a circulation of an electric field that induces magnetic dipole moments in the cylinders with opposite directions between nearest neighbours. The second case will be that of a magnetic field circulation inducing an electric dipole-like moment that, as in the previous case, has opposite directions between nearest neighbours. 
The two modes can be classified as quasi-orthogonal modes that do not interact, as the overlap integrals of their near field distributions are almost vanishing.

By applying the same out-of-plane perturbation as in the case for an isolated particle shown in Fig.~\ref{fig:Figure 1}(b) to each cylinder in the array, omega-like behaviour can be induced. The drilled hole has depth $h_{\rm p}$ and normalized diameter $\sigma_{\rm p}=d_{\rm p}/D_{\rm i}$, where $D_{\rm i}$ is the diameter of a single cylinder and takes the value \(D_{\rm a}\) for the large cylinders and \(D_{\rm b}\) for the small ones. The normalized collective polarizabilities of the metasurface holding q-BICs with this perturbation are shown in Fig.~\ref{fig:Figure 1}(d). It can be clearly seen that the small perforation has introduced $|\hat{\alpha}^{\rm em}|$ and that the electric and magnetic collective polarizabilities have also been perturbed. At the resonant frequency of the collective magnetic polarizability, the collective electric polarizability is affected, and conversely, at the resonance frequency of the collective electric polarizability. This mixing of behaviours and electromagnetic coupling between dipole moments is also visible in the field distributions of the eigenmodes of the out-of-plane asymmetric metasurface, shown in Fig.~\ref{fig:Figure 1}(e). Both modes will present a magnetic dipole and an electric dipole-like field distribution. This is often referred to as mode hybridization~\cite{Gladyshev2020}. The broken out-of-plane symmetry makes it impossible to classify the modes as symmetric and antisymmetric; both behaviours are going to be present. The overlap integrals between the hybridized modes are nonzero, making them interact and exchange energy when they have close resonant wavelengths to each other.

\section{Strong mode coupling via quasi-bound states in the continuum}

At microwave or radio wavelengths, metals do behave as perfect conductors, which makes it possible to achieve very high confinements and avoid strong nonlocal couplings. This facilitates the modeling of the individual polarizabilities of metal elements using analytical expressions, which can then be used to calculate the collective polarizabilities of periodic structures and solve the scattering problem in terms of geometrical parameters. However, at optical wavelengths, mode confinement is much weaker. Even with the use of metals, losses and finite permittivities bound the confinement of the electric field in typical metasurfaces. Furthermore, the use of dielectric materials as substrates, such as silicon oxide, further reduces the possible field confinement granted by a lower index contrast between the structure and the host material or environment. Moreover, dielectric meta-atoms are typically larger, and the ratio between the periodicity of the structure and the wavelength decreases, enhancing nonlocal phenomena such as spatial dispersion.  Due to these aspects, although modeling metasurfaces through collective polarizabilities or susceptibilities provides valuable qualitative insight, it becomes insufficient in the optical domain to provide clear insight of the actual structures required to produce the desired scattering response. 

To overcome these issues, alternative analytical methods are typically employed to elucidate the physical mechanism at play. For this reason, temporal coupled-mode theory (TCMT) has become a powerful framework to analyze the response of resonant nanostructures~\cite{haus2002coupled,CMT-OG}, as it relates the scattering properties of the system to the properties of its eigenmodes. Over the last decades, this formalism has been extensively developed in photonics~\cite{Haus1991,Fan2002,Alpeggiani2017,Zhou2021,Christopoulos2024} and successfully applied to describe a wide variety of photonic systems, including metasurfaces supporting quasi-bound states in the continuum. In this section, we introduce a simple yet useful approach to model omega bianisotropy in metasurfaces using TCMT.

\begin{figure*}[t]
    \centering
    \includegraphics[width=1\linewidth]{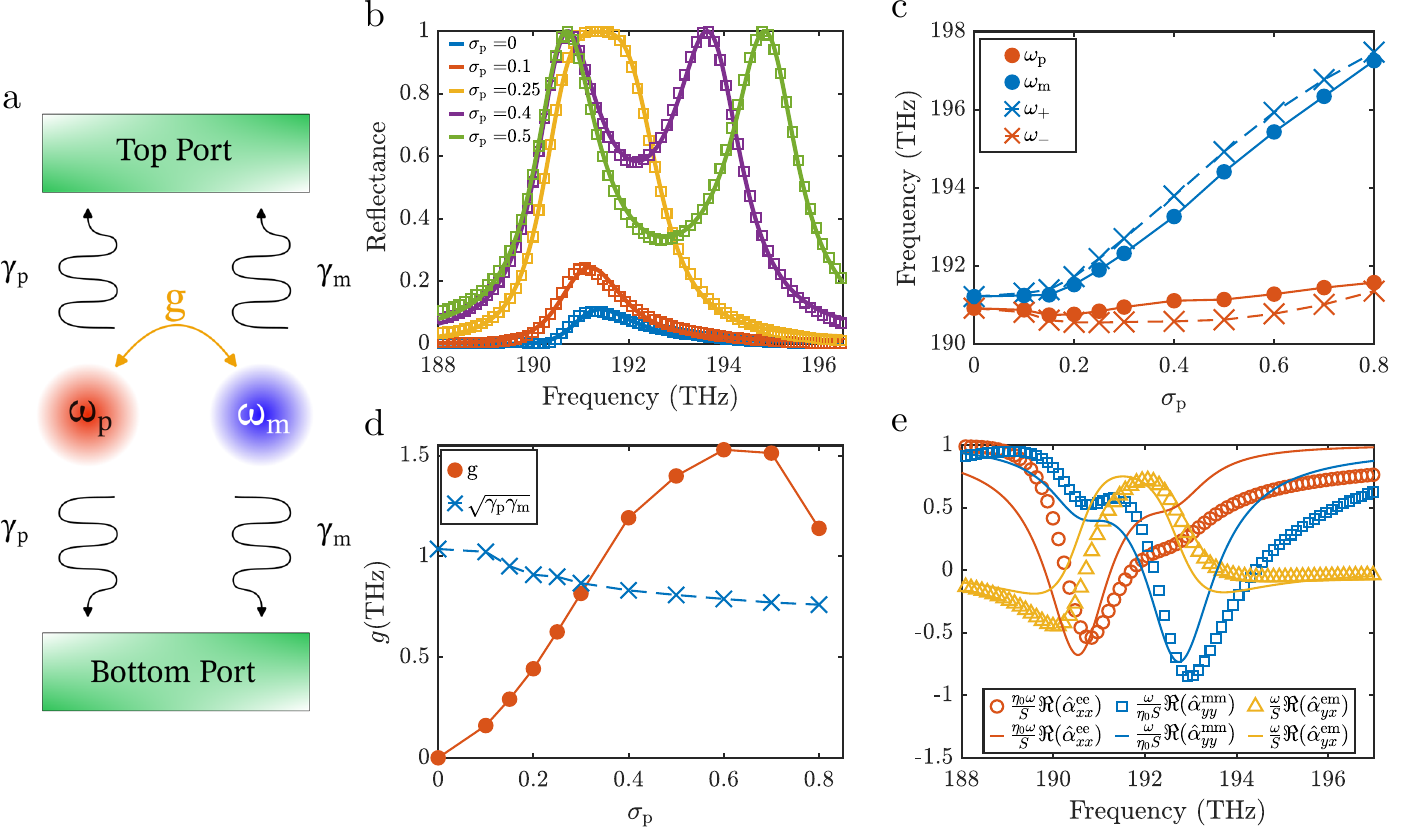}
    \caption{\textbf{Modelling of off-diagonal chiral bianisotropic metasurfaces with TCMT.} (a) Schematic representation of the TCMT model. (b) Comparison between full-wave and TCMT calculations of the reflectance of the metasurface for different values of the normalized diameter $\sigma_{\rm p}=d_{\rm p}/D_{\rm i}$ when $P=1200$ nm, $D=530$ nm, $h=263$ nm,  $\Delta=95$ nm, $h_{\rm p}=20$ nm, and $n=3.5$. For $\sigma_{\rm p}=0$ the system is close to Huygens' condition with $R$ close to zero at the resonant frequency. (c) Resonant frequency of the two hybridized modes for different values of the normalized diameter $\sigma_{\rm p}=d_{\rm p}/D_{\rm i}$. The values for the fitted parameters of the unperturbed structrure are also depicted, $\omega_{\rm p}$ and $\omega_{\rm m}$. (d) Values for the coupling strength, $g$, and the critical condition $g_{\rm c}=\sqrt{\gamma_{\rm p}\gamma_{\rm m}}$ in terms of the normalized diameter. (e) Real part of the collective electric, magnetic, and magnetoelectric polarizabilities as functions of the TCMT parameters, compared with the corresponding values obtained from FEM simulations.}
    \label{fig:Figure 2}
\end{figure*}

Inspired by the analysis made in Section~\ref{sec:Bianisotropy and Hybridization}, we define the unperturbed system supporting two resonant modes: one that decays symmetrically- as an in-plane electric dipole- and one that decays antisymmetrically- as an in-plane magnetic dipole. A schematic of the model is shown in Fig.~\ref{fig:Figure 2}(a). 
The resonant structure is assumed to have uniaxial symmetry, i.e., no polarization conversion is allowed, and thus, it can be modeled with only two ports in the case of normal incidence illumination with no diffraction orders higher than the zeroth being open. Therefore, input waves with vector $|\mathbf{s}_+\rangle=(s_{\rm t +},s_{\rm b +})^T$ and output waves with vector $|\mathbf{s}_-\rangle=(s_{\rm t -},s_{\rm b -})^T$ will only have two elements: a top ($\rm t $) port and a bottom ($\rm b $) port. The unperturbed resonant modes are defined by two eigenfrequencies, labelled as $\omega_p$ and $\omega_m$ for the electric and magnetic resonant modes, respectively.  The decay rates that describe the total modal energy leakage are defined by $\gamma_p$ and $\gamma_m$, for the electric and magnetic resonant modes. Imposing reciprocity on the system~\cite{CMT-OG, CMT-eqs}, the governing equations can be written, denoting $j=\sqrt{-1}$, as
\begin{align}
\frac{d\mathbf{a}}{dt} = \left( j{{\Omega}} - {{\Gamma}} \right)\cdot \mathbf{a} + {{D}}^\mathrm{T} \cdot|\mathbf{s}_+\rangle,\quad
|\mathbf{s}_-\rangle = {{C}}\cdot |\mathbf{s}_+\rangle + {{D}}\cdot \mathbf{a},
\label{eq:CMT}\end{align}
where the amplitudes of the resonances will be contained in the vector $\mathbf{a}=(a_{\rm p},a_{\rm m})$, $\rm p$ and $\rm m$ labels refer to electric and magnetic, and the resonant properties of the structure are defined in the $\Omega$ and $\Gamma$ matrices as
\begin{equation}
    {{\Omega}}=\begin{pmatrix}
        \omega_{\rm p} & g\\
        g & \omega_{\rm m}
    \end{pmatrix},\;
     {{\Gamma}}=\begin{pmatrix}
        \gamma_{\rm p} & 0\\
        0 & \gamma_{\rm m}
    \end{pmatrix}.\;
\end{equation}
Notice that both resonances are coupled by a coupling parameter, $g$, which accounts for the strength of coupling between modes. i.e., the energy exchange ratio between modes~\cite{BIASFREE} that depends on the out-of-diagonal symmetry breaking. 

The $D$ matrix represents the coupling between modes and ports. The $C$ matrix models the direct coupling (direct process or background process) between input and output waves. They are defined as 
\begin{equation}
        {{D}}=
    \begin{pmatrix}
        d_{\rm tp} & d_{\rm tm}\\
        d_{\rm bp} & d_{\rm bm}
    \end{pmatrix},
    \; C
        =e^{j\phi_{\rm}}\begin{pmatrix}
        r & jt_{\rm d}\\
        jt_{\rm d} & r
    \end{pmatrix}.
\end{equation}
    
The elements of the coupling matrix $D$, due to the symmetrical and antisymmetrical nature of the unperturbed resonances, will follow $d_{\rm t p}=d_{\rm b p}$ and $d_{\rm t m}=-d_{\rm b m}$, where $t$ and $b$ refer to the top and bottom port and $p$ and $m$ to the nature of the resonant mode. Lastly, the background process, or direct pathway, matrix $C$, is modeled with the parameters $r_{\rm d}$  and $t_{\rm d}=\sqrt{1-r^2}$  for the module of the reflection and transmission coefficients and $\phi$ for the global phase, which depends on the reference plane used~\cite{Wang2013}.

With the knowledge of the parameters defined previously (eigenfrequencies of the unperturbed modes, decay rates, coupling strength, and reflection coefficient of the background process), scattering parameters of the metasurface can be calculated by solving Eq.~\eqref{eq:CMT} assuming a $e^{j\omega t}$ time dependence, where $\omega$ is the angular frequency. The results for the scattering coefficients are

\begin{equation}\label{eq:Results_scattering_1}
r_{\pm}(\omega)=e^{j\phi}\Bigg(r-\frac{\gamma_{\rm m}A_{\rm p}(\omega)e^{-j\cos^{-1}{(r)}}+\gamma_{\rm p}A_{\rm m}(\omega)e^{j\cos^{-1}{(r)}}\pm 2jg\sqrt{\gamma_{\rm p}\gamma_{\rm m}}}{A_{\rm p}(\omega) A_{\rm m}(\omega)+g^2}\Bigg),
\end{equation}
\begin{equation}\label{eq:Results_scattering_2}
t(\omega)=e^{j\phi}\Bigg(jt_{\rm d}+\frac{\gamma_{\rm m}A_{\rm p}(\omega)e^{-j\cos^{-1}{(r)}}-\gamma_{\rm p}A_{\rm m}(\omega)e^{j\cos^{-1}{(r)}}}{A_{\rm p}(\omega) A_{\rm m}(\omega)+g^2}\Bigg),
\end{equation}

with functions of the angular frequency defined as $ A_{\rm p}(\omega)=j(\omega-\omega_{\rm p})+\gamma_{\rm p}$ for the electric resonance and $A_{\rm m}(\omega)=j(\omega-\omega_{\rm m})+\gamma_{\rm m}$ for the magnetic resonance. 
Due to the coupling between resonances, the scattering matrix of the process will turn into
\begin{equation}
    |\mathbf{s}_-\rangle=\begin{pmatrix}
        r_{-} & t\\
        t & r_+
    \end{pmatrix}|\mathbf{s}_+\rangle,
\end{equation}
with $|r_+|=|r_-|$. 

To check the validity of the TCMT model, we perform full-wave calculations using the structure presented in the previous section and fit the parameters to check the role that they play. The simulated structure, which has $C_{\rm 4v}$ symmetry, shares the same form of $S$ matrix. The reflectance for several values of $\sigma_{\rm p}$ is shown in Fig.~\ref{fig:Figure 2}(b). The obtained values via FEM simulations are plotted as squares, while the TCMT fit is plotted as a solid line, showing good agreement. Additionally, the retrieved values for the eigenfrequencies from the fit are shown in Fig.~\ref{fig:Figure 2}(c), and for the coupling strength in Fig.~\ref{fig:Figure 2}(d). 

When mirror symmetry is present, $\sigma_{\rm p}=0$, the structure is in a Huygens' configuration, $\omega_{\rm p}\simeq\omega_{\rm m}$. A pair of even and odd resonances, in this case, electric and magnetic dipole q-BIC resonances, overlap, resulting in transmission close to one. When out-of-plane symmetry is broken, electromagnetic coupling is enabled, and the modes hybridize. The coupling parameter, $g$, effectively models this electromagnetic omega coupling or, equivalently, the hybridization of the modes. Up to a point, increasing the diameter of the mirror-breaking perturbation enhances the mode overlap, thereby increasing the coupling strength between the resonances. In the scattering parameters, this effect is translated into an increase in reflectance. However, in detail, two distinct effects are taking place. Firstly, perforating the top of the structure results in a bigger shift in frequency for the unperturbed magnetic dipole resonance than for the unperturbed electric dipole resonance. The unperturbed magnetic dipole resonance is a product of the circulation of the electric displacement field, which is highly dependent on the height of the resonator. This is shown in Fig.~\ref{fig:Figure 2}(c) for the values of $\omega_{\rm p}$ and $\omega_{\rm m}$. Secondly, in Fig.~\ref{fig:Figure 2}(d), it can be seen how the coupling strength increases with the value of $\sigma_{\rm p}$ until it reaches its maximum near $\sigma_{\rm p}=\sqrt{2}/2\simeq0.707$, and then decreases. This can be explained as for $\sigma_{\rm p}=\sqrt{2}/2$ the area of the perforation is exactly half of the area of the cross-section of the cylinder.

Indeed, with this formulation of the TCMT model, an interesting property of coupled systems arises. The new eigenfrequencies of the perturbed coupled system, corresponding to the hybridized modes, can be calculated in terms of the eigenfrequencies of the uncoupled (unperturbed) system and the coupling strength as
\begin{equation}
    \omega_{\pm}=\left(\frac{\omega_{\rm p}+\omega_{\rm m}}{2}\right)\pm \sqrt{ \left(\frac{\omega_{\rm p}-\omega_{\rm m}}{2}\right)^2+g^2},
\end{equation}
where $\omega_{+}$ and $\omega_{-}$ are the high-energy hybrid mode (HEHM) and the low-energy hybrid mode (LEHM) resonance frequencies, respectively. This phenomenon is known to occur in two-level systems, between excitonic resonances and optical resonances~\cite{Dovzhenko2018,vandeGroep2013,Xie2021,Qin2023,Weber2023}, as well as between optical modes~\cite{Fan2025,Kumar2025,Deng2025}. As $g$ increases, the two branches separate further from each other, and an anticrossing behaviour appears. This anticrossing nature is usually referred to as Rabi splitting due to the similarity of this effect and the strong interaction between two quantum states~\cite{Santhosh2016,Reithmaier2004}. In Fig.~\ref{fig:Figure 2}(c), the two branches are plotted for completeness.

Figure~\ref{fig:Figure 2}(d) shows the obtained values for the coupling strength in terms of $\sigma_{\rm p}$ and the value $g_{\rm c}=\sqrt{\gamma_{\rm p}\gamma_{\rm m}}$. For the critical value of $g_{\rm c} $, for the specific case of $\omega_{\rm p}=\omega_{\rm m}=\omega_0$, the polarizabilities correspond to a balanced particle case, i.e. at $\omega=\omega_0$,  $\alpha^{\rm ee}\eta_0=\alpha^{\rm mm}/\eta_0=\alpha^{\rm em}$. In Fig.~\ref{fig:Figure 2}(c), it is shown that for values greater than $\sigma_{\rm p}=0.3$, the coupling strength exceeds this critical value. In the literature, it is also possible to find the value defined to discern between a weak coupling regime and a strong coupling when $\omega_{\rm p}=\omega_{\rm m}$ as the sum of the decay rates (widths of the resonances) divided by two~\cite{Liu2014,Ibrahim2021}. This value is chosen because it is similar to the Rayleigh criterion for angular resolution of optical systems. It is considered that such a coupling strength will separate the branches sufficiently to be differentiated, forming an anticrossing pattern that allows them to be well distinguished. Using that condition, $g>\sqrt{(\gamma_{\rm p}^2+\gamma_{\rm m}^2)/2}=g_{\rm strong}$ is obtained as the strong coupling condition. 

To further demonstrate that the TCMT explains the physics behind the structure, Fig.~\ref{fig:Figure 2}(e) shows the real part of the collective polarizabilities as functions of the TCMT parameters, compared to the values obtained from the FEM simulations. These quantities, which depend on the real and imaginary parts of the scattering coefficients, show good agreement between simulations and the TCMT. The discrepancy is present due to the quality factor of the resonances, which is in the order of the hundreds; TCMT works better for higher quality factors. Furthermore, the real background process is not constant.
The expressions for the collective polarizabilities in terms of scattering parameters can be written as~\cite{albooyeh2023classification}

\begin{align}\label{eq: Pol_ee}
    \hat\alpha_{xx}^{\rm ee}\eta_0=\hat\alpha_{yy}^{\rm ee}\eta_0=j\dfrac{S}{2\omega}(r_+-r_-+2t-2),\\
    \hat\alpha_{xx}^{\rm mm}/\eta_0=\hat\alpha_{yy}^{\rm mm}/\eta_0=j\dfrac{S}{2\omega}(2t-2-r_+-r_-),\\
    \hat\alpha_{xy}^{\rm em}=-\hat\alpha_{yx}^{\rm em}=-\hat\alpha_{yx}^{\rm me}=\hat\alpha_{xy}^{\rm me}=-j\dfrac{S}{2\omega}(r_+-r_-),\label{eq: Pol_em}
\end{align}
where $S$ is the area of the meta-atom. Taking into account the results of the TCMT, Eqs.~\eqref{eq:Results_scattering_1}-\eqref{eq:Results_scattering_2}, these quantities can be written in the next form

\begin{equation}\label{eq: CMT pol ee}
    \dfrac{\hat{\alpha}_{xx}^{\rm ee}}{\eta_0}=\frac{je^{j\phi}}{\omega}\Bigg(jt_{\rm d}+r-2\frac{\gamma_{\rm p}A_{\rm m}(\omega)e^{j\cos^{-1}(r)}}{(A_{\rm p}(\omega)A_{\rm m}(\omega)+g^2)}\Bigg)-j/\omega,
\end{equation}

\begin{equation}\label{eq: CMT pol mm}
            \dfrac{\hat{\alpha}_{yy}^{\rm mm}}{\eta_0}=\frac{je^{j\phi}}{\omega}\Bigg(jt_{\rm d}-r+2\frac{\gamma_{\rm m}A_{\rm p}(\omega)e^{-j\cos^{-1}(r)}}{A_{\rm p}(\omega)A_{\rm m}(\omega)+g^2}\Bigg)-j/\omega,
\end{equation}
\begin{equation}\label{eq: CMT pol em}
    \hat{\alpha}_{xy}^{\rm em}=-\frac{e^{j\phi}2g\sqrt{\gamma_{\rm p}\gamma_{\rm m}}}{A_{\rm p}(\omega)A_{\rm m}(\omega)+g^2\omega},
\end{equation}
where it can be seen how, in the first order approximation, $\hat{\alpha}_{xy}^{\rm em}$ is linearly proportional to $g$, while $\hat{\alpha}_{xx}^{\rm ee}$ and $\hat{\alpha}_{yy}^{\rm mm}$ are not affected.

\begin{figure*}[t]
    \centering
    \includegraphics[width=1\linewidth]{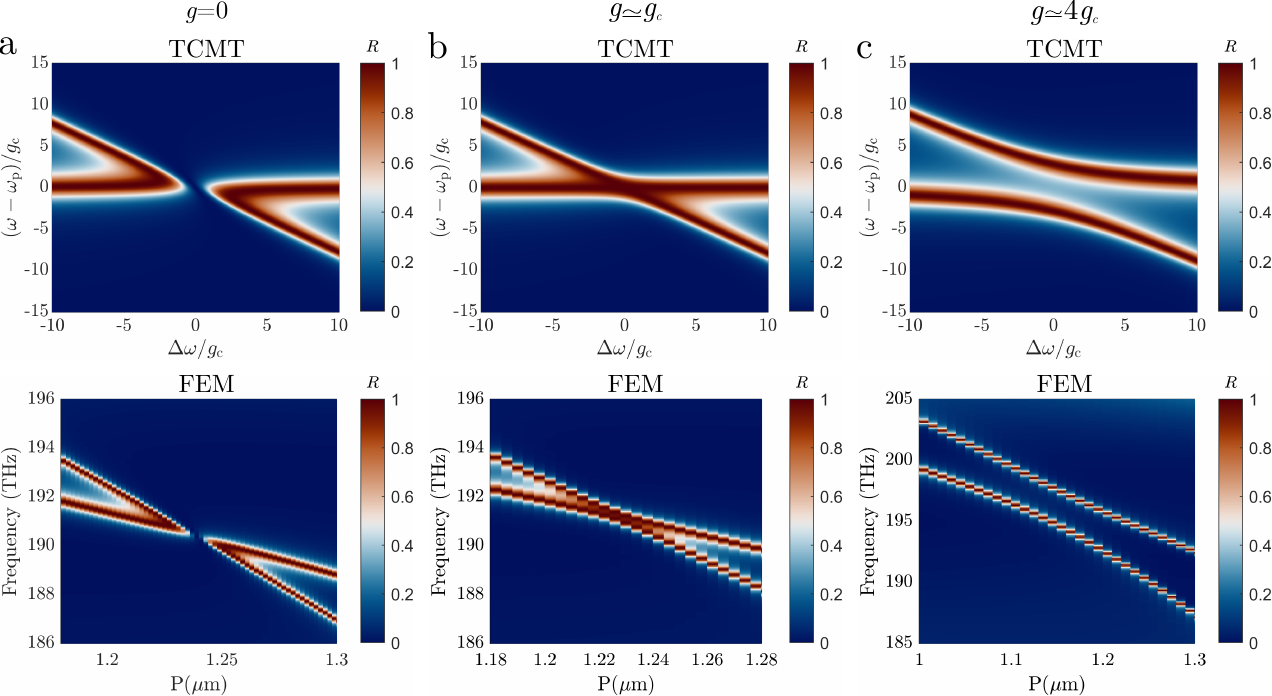}
    \caption{\textbf{Scattering properties of resonance crossings in off-diagonal chiral type bianisotropic metasurfaces for different coupling scenarios.} TCMT results have decay rates chosen as $\gamma_{\rm m}=3\gamma_{\rm p}$, with constant background process $r=0$, and the horizontal axis is normalized as $\Delta \omega=\omega_{\rm p}-\omega_{\rm m}$. (a) $\sigma_{\rm p}=0$ and $g=0$. No coupling between electric dipole resonance and magnetic dipole resonance, as $\sigma_{\rm h}$ is a symmetry. At $\Delta \omega=0$, a Huygens-like pair is created, achieving $R=0$ and transmittance $T=1$. (b) $g\simeq g_{\rm c}$ and $\sigma_{\rm p}=0.12$, at the crossing point, full reflection is achieved. (c) Strong coupling regime, $\sigma_{\rm p}=0.5$ and $g\simeq4g_{\rm c}$, at the same time $g>g_{\rm strong}$. At $\Delta \omega=0$, a clear anticrossing can be observed. The Rabi-like splitting can be controlled with the perturbation, as shown in Fig.~2. Geometrical parameters for the metasurface: $D=530$ nm, $\Delta=40$ nm, $h=263$ nm, $h_{\rm p}=20$ nm.}
    \label{fig:Figure 3}
\end{figure*}

In terms of the coupling strength and the scattering properties, there are three cases predicted by this TCMT model that have special interest. In Fig.~\ref{fig:Figure 3}, the three scenarios are shown, each one showing the TCMT prediction versus the FEM simulated structure. In all FEM plots, the periodicity of the structure is tuned, making the eigenfrequencies of the electric and magnetic dipole-like resonances cross. The uncoupled electric dipole resonance is altered very slightly with the change of $P$, whereas the uncoupled magnetic dipole resonance shifts rapidly towards lower frequencies. Fig.~\ref{fig:Figure 3}(a) depicts the crossing of both resonances in the case where no coupling is allowed, which is the same scenario as in ref.~\cite{CMT-OG}. Fig.~\ref{fig:Figure 3}(b) shows the case in which $g\simeq g_{\rm c}=\sqrt{\gamma_{\rm p}\gamma_{\rm m}}$. It can be seen how, at the crossing, maximum reflection is achieved for a perfectly reflecting band broader than each resonance isolated. Lastly, in Fig.~\ref{fig:Figure 3}(c), the anticrossing feature resulting from the strong interaction of the modes is shown. Here, the system is in the known strong coupling regime. A necessary remark is that all these results are calculated for the structure in Fig.~\ref{fig:Figure 1}(d), therefore, the background process has $r\rightarrow0$. 

\section{Dual-band asymmetric absorption mediated by strongly coupled modes}

In this section, building upon the preceding discussion, we present a general theoretical framework for the design of reciprocal bianisotropic metasurfaces that achieve maximum difference in directional absorption at optical frequencies. All discussion up to this point has assumed a passive, non-lossy character for the materials that comprise the structure. However, introducing losses to the broken $\sigma_{\rm h}$ structure opens up the opportunity to have $|r^-|^2\neq|r^+|^2$. Indeed, with the addition of losses, asymmetric absorption is possible as $\Delta A=A^+-A^-=|r^-|^2-|r^+|^2$, where $A^{+}$ is the absorption calculated for a wave with $k_z>0$ and $A^-$ is the absorption calculated for a wave with $k_z<0$. Losses can be added into the system as a perturbation when $\gamma<<\omega$, separating $\gamma_{\rm i} \rightarrow\gamma_{\rm i}^{\rm r}+\gamma_{\rm i}^{\rm l}$ where $\gamma_{\rm i}^{\rm r}$ is the radiative part of the decay rate and $\gamma_{\rm i}^{\rm l}$ is the part related to intrinsic losses of the materials. 

For the simplest case in which $\omega_{\rm p}=\omega_{\rm m}$ and $\gamma=\gamma_{\rm p}^{\rm r}=\gamma_{\rm p}^{\rm l}=\gamma_{\rm m}^{\rm r}=\gamma_{\rm m}^{\rm l}$, the difference in absorption between sides, evaluated at the LEHM and HEHM frequencies, takes the form of 
\begin{equation}\label{eq: Delta Absorption}
    \Delta A|_{\omega=\omega_{\pm}}=\mp \frac{g^2r}{\gamma^2 + g^2}.
\end{equation}
This equation shows that asymmetric absorption in the system is bounded by the reflection coefficient of the direct process. This is similar to reported findings in~\cite{Wang2013}, where the background process was found to bound the asymmetric decay of a single resonance system. In addition, following Eq.~\eqref{eq: Delta Absorption}, it can be seen that in order to maximize $\Delta A$ at the resonant frequency of the hybrid mode, it is necessary to have a background process with a high reflection coefficient. 

\begin{figure*}[t]
    \centering
    \includegraphics[width=1\linewidth]{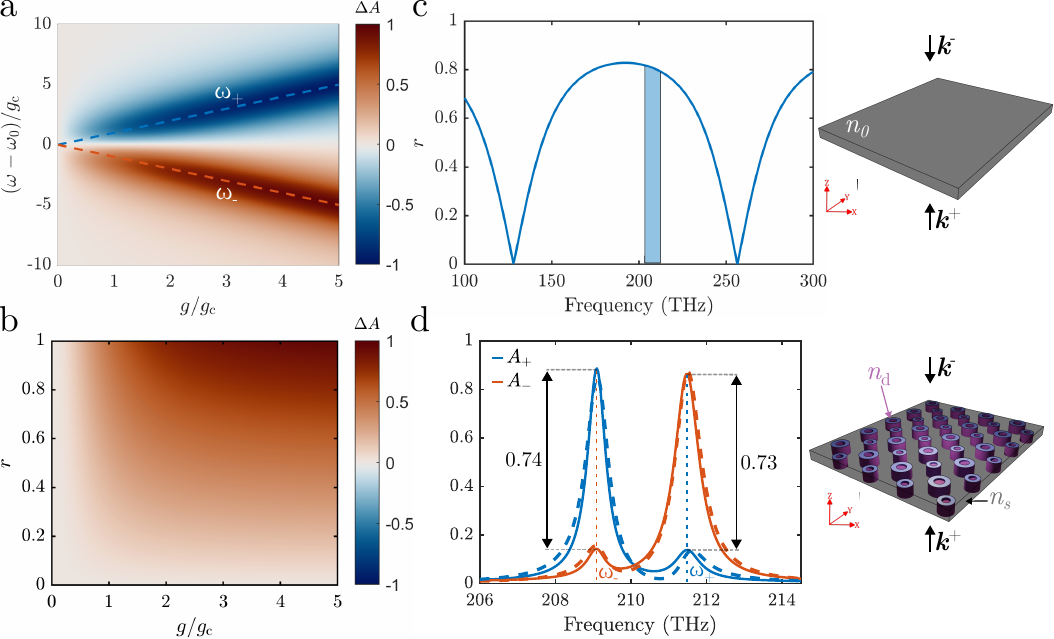}
    \caption{\textbf{Dual-band absorption in strongly coupled metasurfaces with off-diagonal chirality or omega-type bianisotropy.} (a) Case in which $\omega_{\rm p}=\omega_{\rm m}=\omega_{0}$, $\gamma=\gamma_{\rm p}^{\rm r}=\gamma_{\rm p}^{\rm l}=\gamma_{\rm m}^{\rm r}=\gamma_{\rm m}^{\rm l}$, and $r=1$. The difference in absorption grows with the coupling strength for each branch of the hybrid modes. (b) Absorption difference evaluated at $\omega_{+}$ for $\omega_{\rm p}=\omega_{\rm m}=\omega_{0}$, $\gamma=\gamma_{\rm p}^{\rm r}=\gamma_{\rm p}^{\rm l}=\gamma_{\rm m}^{\rm r}=\gamma_{\rm m}^{\rm l}$ in terms of absolute value of the reflection coefficient for the background process and coupling strength. (c) Reflection from a Fabry-Pérot slab of dielectric of height $h=360$ nm and effective refractive index $n_{0}=(n_{\rm s}+n_{\rm d})/2=3.25$ . (d) Calculated absorption via FEM simulations for the metasurface embedded in a dielectric slab. Solid lines represent FEM results, while the TCMT fit is represented as dashed lines. Absorption FEM geometrical parameters: $n_{\rm s}=2.5$, $n_{d}=4$, $P=510$ nm, $\Delta=70$ nm, $h_{\rm p}=40$ nm, $\sigma_{\rm p}=0.5$, $h=360$ nm, $D=280$ nm, and $k_{\rm}=0.005$ as the imaginary part of the refractive index.}
    \label{fig:Figure 4}
\end{figure*}

To further elucidate the results, we show in Fig.~\ref{fig:Figure 4}(a) the dependence of the absorption difference with the coupling coefficient for the case in which $\omega_{\rm m}=\omega_{\rm p}=\omega_0$, $r=1$, and $\gamma=\gamma_{\rm p}^{\rm r}=\gamma_{\rm p}^{\rm l}=\gamma_{\rm m}^{\rm r}=\gamma_{\rm m}^{\rm l}$. The axes are normalized using $g_{\rm c}=\sqrt{\gamma_{\rm p}^{\rm r}\gamma_{\rm m}^{\rm r}}$. It can be seen how for each branch $\omega_+$ and $\omega_-$, the difference in absorption increases with bigger values of $g$. Interestingly, at a given value of $g$, in the frequency spectrum around $\omega_0$ two distinct bands will rise. At lower frequencies, one side of the structure will have almost full absorption around $\omega\simeq\omega_-$ while the other will have full reflection. At higher frequencies, $\omega\simeq\omega_+$, the responses are going to be interchanged. It is worth noting that the structure is reciprocal; therefore, for $\Delta A=\pm 1$, the transmission vanishes, i.e., $|t|^2=0$.

Fig.~\ref{fig:Figure 4}(b) shows the dependence of $\Delta A$ on the coupling strength and reflection coefficient, as Eq.~\eqref{eq: Delta Absorption} predicts. Following the recipe provided by the equation, we demonstrate how to design a metasurface with a similar scattering response, i.e., with maximum directional absorption difference. The main idea is to embed the above shown metasurface, with broken out-of-plane symmetry, in a dielectric slab. The slab will emulate a direct scattering process with high reflectivity, specifically a Fabry-Pérot resonator. We show in Fig.~\ref{fig:Figure 4}(c) the broadband response of a slab with effective refractive index $n_{0}=(n_{\rm s}+n_{\rm d})/2$, where $n_{\rm s}$ is the refractive index of the final slab and $n_{\rm d}$ is the refractive index of the structure that will be embbeded, similar proceeding than that used in~\cite{Fan2002}. Due to the fact that the decay rates of the Fabry-Pérot resonances of the slab are significantly larger than those of the resonances held by the metasurface, it can be regarded as a constant direct pathway. We choose a thickness so that the module of the reflectivity is high near 210 THz, $r\simeq0.8$, as shown in the blue region of Fig.~\ref{fig:Figure 4}(c). Afterwards, the metasurface geometrical parameters are chosen so that $\omega_{\rm m}=\omega_{\rm p}$. The drilled hole on top of the nanodisks will enable strong mode coupling between the modes and bianisotropic omega behaviour. 

The result from the FEM simulations is shown in Fig.~\ref{fig:Figure 4}(d), where the dual band behaviour can be clearly seen. The symmetry with respect to 210 THz in the response is also large due to $\gamma_{\rm p}\simeq\gamma_{\rm m}$. Fitted results using the TCMT are plotted as dashed lines showing good agreement. With this straightforward approach, and without any optimization, more than a $70\%$ difference in directional absorption is reached at the resonant frequencies of the HEHM and the LEHM. The calculated coupling strength is equal to $g=1.21$ THz, which is 5.72 times larger than $g_{\rm strong}=0.21$ THz; therefore, the modes are strongly coupled. Nevertheless, the modes do not have the same scattering decay rates. However, a rough estimation of the limit can still be performed using Eq.~\eqref{eq: Delta Absorption}. Plugging $g$, $\alpha=0.21$ THz, and the fitted reflectivity for the background, $r=0.8$, the result is that $\Delta A|_{\omega=\omega_{\pm}}=\mp0.78$, which is a value near the one obtained with FEM simulations at $\omega_+$ and $\omega_-$ for the difference in directional absorption. The presented design exemplifies the practical implementation of the model, serving as a theoretical approach for designing metasurface-based devices that require directional control of light absorption. To maximize $\Delta A$, multiple strategies could be further implemented: improving the background to have higher reflectivity, using higher quality factor resonances, or further increasing $g$ by using a deeper perforation on top of the cylinders.

\section{Conclusions}
In this work, we have developed a comprehensive and physically intuitive framework to describe omega-type bianisotropy in dielectric metasurfaces operating under high-quality resonances. By using a temporal coupled-mode theory (TCMT) model, we have shown how controlled out-of-plane symmetry breaking activates electromagnetic coupling between electric and magnetic dipolar resonances of opposite parity, leading to their hybridization and to the emergence of a Rabi-like splitting characteristic of the strong coupling regime. The coupling can be controlled by the amount of out-of-plane symmetry breaking introduced and can lead to a very strong bianisotropic response in a deeply subwavelength structure. In fact, the structure shown is a reciprocal analogue of a pure moving medium as that shown in~\cite{BIASFREE}. In contrast to many works in the literature, the anticrossing behavior of the system can be observed directly from the parameters calculated via TCMT, rather than as the solutions of an effective Hamiltonian and subsequent Fano fitting. 
 
 Furthermore, this study presents a comprehensive theoretical model for designing optical reciprocal metasurfaces with maximum asymmetric directional absorption. We have demonstrated that when intrinsic losses are introduced, these strongly coupled hybrid modes produce a dual-band asymmetric absorption response under counter-propagating illumination. A realization of such an optical response can be achieved by embedding the metasurface in a carefully designed dielectric slab acting as a Fabry–Pérot resonator, which enhances the background reflectivity, allowing the directional absorption contrast to exceed $70\%$ in two distinct spectral bands. This behavior represents an efficient and fully reciprocal approach to implementing maximum direction-dependent light dissipation in passive optical systems.

\section*{Acknowledgement}
LMME and ADR acknowledge the Spanish National Research Council (grant No. PID2021-128442NA-I00 and CNS2024-154715). LMME acknowledges Universitat Politècnica de València (PAID-01-23). BA and VA acknowledge the Research Council of Finland grant no. 356797 and Research Council of Finland Flagship Program (Grant No.: 320167, PREIN). VA acknowledges the 
Finnish Foundation for Technology Promotion.

\section*{Data Availability Statement}
The data that support the findings of this study are available from the corresponding author upon reasonable request.

\nolinenumbers
\printbibliography

@article{albooyeh2023classification,
  title={Classification of bianisotropic metasurfaces from reflectance and transmittance measurements},
  author={Albooyeh, M and Asadchy, V and Zeng, J and Rajaee, M and Kazemi, H and Hanifeh, M and Capolino, F},
  journal={ACS Photonics},
  volume={10},
  number={1},
  pages={71--83},
  year={2023},
  publisher={ACS Publications}
}

@article{alaee2018electromagnetic,
  title={An electromagnetic multipole expansion beyond the long-wavelength approximation},
  author={Alaee, Rasoul and Rockstuhl, Carsten and Fernandez-Corbaton, Ivan},
  journal={Optics Communications},
  volume={407},
  pages={17--21},
  year={2018},
  publisher={Elsevier}
}

@article{Liu2014,
   author = {Xiaoze Liu and Tal Galfsky and Zheng Sun and Fengnian Xia and Erh Chen Lin and Yi Hsien Lee and Stéphane Kéna-Cohen and Vinod M. Menon},
   doi = {10.1038/nphoton.2014.304},
      
   issue = {1},
   journal = {Nature Photonics},
    
   pages = {30-34},
   publisher = {Nature Publishing Group},
   title = {Strong light-matter coupling in two-dimensional atomic crystals},
   volume = {9},
   year = {2014},
}

@article{Ibrahim2021,
   author = {Ibrahim A.M. Al-Ani and Khalil As'Ham and Lujun Huang and Andrey E. Miroshnichenko and Haroldo T. Hattori},
      
   issue = {12},
   journal = {Laser and Photonics Reviews},
    
   publisher = {John Wiley and Sons Inc},
   title = {Enhanced Strong Coupling of TMDC Monolayers by Bound State in the Continuum},
   volume = {15},
   year = {2021},
}

@article{Wang2013,
author = {Ken Xingze Wang and Zongfu Yu and Sunil Sandhu and Shanhui Fan},
journal = {Opt. Lett.},
number = {2},
pages = {100--102},
publisher = {Optica Publishing Group},
title = {Fundamental bounds on decay rates in asymmetric single-mode optical resonators},
volume = {38},
year = {2013},
doi = {10.1364/OL.38.000100},
}

@article{Hu2020,
   author = {Jack Hu and Mark Lawrence and Jennifer A. Dionne},
   doi = {10.1021/acsphotonics.9b01352},
      
   issue = {1},
   journal = {ACS Photonics},
    
   pages = {36-42},
   publisher = {American Chemical Society},
   title = {High Quality Factor Dielectric Metasurfaces for Ultraviolet Circular Dichroism Spectroscopy},
   volume = {7},
   year = {2020}
}

@article{Geometry-1,
   author = {Mark Lawrence and Jennifer A. Dionne},
      
   issue = {1},
   journal = {Nature Communications},
    
   pmid = {31341164},
   publisher = {Nature Publishing Group},
   title = {Nanoscale nonreciprocity via photon-spin-polarized stimulated Raman scattering},
   volume = {10},
   year = {2019}
}

@article{Haus1991,
  author={Haus, H.A. and Huang, W.},
  journal={Proceedings of the IEEE}, 
  title={Coupled-mode theory}, 
  year={1991},
  volume={79},
  number={10},
  pages={1505-1518},
  doi={10.1109/5.104225}}

@article{Fan2002,
   author = {Shanhui Fan},
   doi = {10.1063/1.1448174},
      
   issue = {6},
   journal = {Applied Physics Letters},
    
   pages = {908-910},
   title = {Sharp asymmetric line shapes in side-coupled waveguide-cavity systems},
   volume = {80},
   year = {2002},
}

@article{Christopoulos2024,
   author = {Thomas Christopoulos and Odysseas Tsilipakos and Emmanouil E. Kriezis},
      
   issue = {1},
   journal = {Journal of Applied Physics},
    
   publisher = {American Institute of Physics},
   title = {Temporal coupled-mode theory in nonlinear resonant photonics: From basic principles to contemporary systems with 2D materials, dispersion, loss, and gain},
   volume = {136},
   year = {2024},
}

@article{CMT-eqs,
   author = {Zhexin Zhao and Cheng Guo and Shanhui Fan},
      
   issue = {3},
   journal = {Physical Review A},
    
   publisher = {American Physical Society},
   title = {Connection of temporal coupled-mode-theory formalisms for a resonant optical system and its time-reversal conjugate},
   volume = {99},
   year = {2019},
}

@article{Gladyshev2020,
  title={Symmetry analysis and multipole classification of eigenmodes in electromagnetic resonators for engineering their optical properties},
  author={Gladyshev, Sergey and Frizyuk, Kristina and Bogdanov, Andrey},
  journal={Physical Review B},
  volume={102},
  number={7},
  pages={075103},
  year={2020},
  publisher={APS}
}

@article{Poleva2023,
   author = {Maria Poleva and Kristina Frizyuk and Kseniia Baryshnikova and Andrey Evlyukhin and Mihail Petrov and Andrey Bogdanov},
      
   issue = {4},
   journal = {Physical Review B},
    
   publisher = {American Physical Society},
   title = {Multipolar theory of bianisotropic response of meta-atoms},
   volume = {107},
   year = {2023}
}

@article{Alpeggiani2017,
   author = {Filippo Alpeggiani and Nikhil Parappurath and Ewold Verhagen and L. Kuipers},
      
   issue = {2},
   journal = {Physical Review X},
   publisher = {American Physical Society},
   title = {Quasinormal-mode expansion of the scattering matrix},
   volume = {7},
   year = {2017}
}

@article{Albooyeh2015,
  title = {Revisiting substrate-induced bianisotropy in metasurfaces},
  author = {Albooyeh, M. and Alaee, R. and Rockstuhl, C. and Simovski, C.},
  journal = {Phys. Rev. B},
  volume = {91},
  issue = {19},
  pages = {195304},
  numpages = {11},
  year = {2015},
  publisher = {American Physical Society},
  doi = {10.1103/PhysRevB.91.195304},
}

@article{BIASFREE,
  title={Giant bias-free nonreciprocity for unpolarized light via synthetic motion},
  author={L. M. Máñez-Espina and B. Amrahi and I. Faniayeu and R. Cichelero and A. Dmitriev and A. Díaz-Rubio and V. S. Asadchy},
  journal={arXiv preprint arXiv:2510.14069},
  year={2025}
}

@article{Santhosh2016,
   author = {Kotni Santhosh and Ora Bitton and Lev Chuntonov and Gilad Haran},
      
   journal = {Nature Communications},
    
   publisher = {Nature Publishing Group},
   title = {Vacuum Rabi splitting in a plasmonic cavity at the single quantum emitter limit},
   volume = {7},
   year = {2016},
}

@article{Reithmaier2004,
author={Reithmaier, J. P.
and S{\k{e}}k, G.
and L{\"o}ffler, A.
and Hofmann, C.
and Kuhn, S.
and Reitzenstein, S.
and Keldysh, L. V.
and Kulakovskii, V. D.
and Reinecke, T. L.
and Forchel, A.},
title={Strong coupling in a single quantum dot--semiconductor microcavity system},
journal={Nature},
year={2004},
day={01},
volume={432},
number={7014},
pages={197-200},
doi={10.1038/nature02969}
}

@article{Huang2024,
  title={Realizing ultrahigh-Q resonances through harnessing symmetry-protected bound states in the continuum},
  author={Huang, Lujun and Li, Shuangli and Zhou, Chaobiao and Zhong, Haozong and You, Shaojun and Li, Lin and Cheng, Ya and Miroshnichenko, Andrey E},
  journal={Advanced Functional Materials},
  volume={34},
  number={11},
  pages={2309982},
  year={2024},
  publisher={Wiley Online Library}
}

@article{Liu2017,
   author = {Yong Chun Liu and Bei Bei Li and Yun Feng Xiao},
   doi = {10.1515/nanoph-2016-0168},
      
   issue = {5},
   journal = {Nanophotonics},
    
   pages = {789-811},
   publisher = {Walter de Gruyter GmbH},
   title = {Electromagnetically induced transparency in optical microcavities},
   volume = {6},
   year = {2017},
}

@article{Deng2025,
   author = {Huachun Deng and Xiong Jiang and Yao Zhang and Yixuan Zeng and Hamdi Barkaoui and Shumin Xiao and Shaohua Yu and Yuri Kivshar and Qinghai Song},
   journal = {Sci. Adv},
   pages = {9562},
   title = {Chiral lasing enabled by strong coupling},
   volume = {11},
   year = {2025},
}

@book{LibroAzul,
title = {Electromagnetics of bi-anisotropic materials: Theory and applications},
author = {Anatoly Serdyukov and Igor Semchenko and Sergei Tretyakov and Ari Sihvola},
year =  {2001},
publisher = {Gordon and Breach Science Publishers},
address = {United States},
}

@article{Zhou2021,
author={Zhou, Ming
and Liu, Dianjing
and Belling, Samuel W.
and Cheng, Haotian
and Kats, Mikhail A.
and Fan, Shanhui
and Povinelli, Michelle L.
and Yu, Zongfu},
title={Inverse Design of Metasurfaces Based on Coupled-Mode Theory and Adjoint Optimization},
journal={ACS Photonics},
year={2021},
  
day={18},
}

@book{Absorption_Scattering_Small_Particles_book,
   author = {Craig F. Bohren and Donald R. Huffman},
    
   publisher = {John Wiley \& Sons},
   title = {Absorption and Scattering of Light by Small Particles},
   year = {2008},
}

@article{Baranov2018,
   author = {Denis G. Baranov and Martin Wersäll and Jorge Cuadra and Tomasz J. Antosiewicz and Timur Shegai},
   doi = {10.1021/acsphotonics.7b00674},
      
   issue = {1},
   journal = {ACS Photonics},
    
   pages = {24-42},
   publisher = {American Chemical Society},
   title = {Novel Nanostructures and Materials for Strong Light-Matter Interactions},
   volume = {5},
   year = {2018},
}

@article{Tretyakov1993,
author = {S.A. Tretyakov  and A.A. Sochava },
title = {Proposed composite material for nonreflecting shields and antenna radomes},
journal = {Electronics Letters},
volume = {29},
issue = {12},
pages = {1048-1049},
year = {1993},
doi = {10.1049/el:19930699}
}

@article{Radi2013,
  author={Ra'di, Younes and Asadchy, Viktar S. and Tretyakov, Sergei A.},
  journal={IEEE Transactions on Antennas and Propagation}, 
  title={Total Absorption of Electromagnetic Waves in Ultimately Thin Layers}, 
  year={2013},
  volume={61},
  number={9},
  pages={4606-4614},
  doi={10.1109/TAP.2013.2271892}}

@article{Yazdi2015,
   author = {Mohammad Yazdi and Mohammad Albooyeh and Rasoul Alaee and Viktar Asadchy and Nader Komjani and Carsten Rockstuhl and Constantin R. Simovski and Sergei Tretyakov},
   doi = {10.1109/TAP.2015.2423855},
      
   issue = {7},
   journal = {IEEE Transactions on Antennas and Propagation},
    
   pages = {3004-3015},
   publisher = {Institute of Electrical and Electronics Engineers Inc.},
   title = {A Bianisotropic Metasurface with Resonant Asymmetric Absorption},
   volume = {63},
   year = {2015},
}

@article{Fan2025,
   author = {Jinying Fan and Zhanqiang Xue and Ye Zhou and Longqing Cong},
      
   journal = {Laser and Photonics Reviews},
   publisher = {John Wiley and Sons Inc},
   title = {Directional Loss Immune Metasurfaces},
   year = {2025},
}

@article{Simovski1997,
author = {C.R. Simovski and S.A. Tretyakov and A.A. Sochava and B. Sauviac and F. Mariotte and T.G. Kharina},
title = {Antenna Model for Conductive Omega Particles},
journal = {Journal of Electromagnetic Waves and Applications},
volume = {11},
number = {11},
pages = {1509--1530},
year = {1997},
publisher = {Taylor \& Francis},
doi = {10.1163/156939397X00567},
}

@article{Asadchy2018,
   author = {Viktar S. Asadchy and Ana Díaz-Rubio and Sergei A. Tretyakov},
   doi = {10.1515/nanoph-2017-0132},
     
   issue = {6},
   journal = {Nanophotonics},
    
   pages = {1069-1094},
   publisher = {Walter de Gruyter GmbH},
   title = {Bianisotropic metasurfaces: Physics and applications},
   volume = {7},
   year = {2018},
}

@article{Kumar2025,
   author = {Brijesh Kumar and Ivan Toftul and Anshuman Kumar and Maxim Gorkunov and Yuri Kivshar},
   journal = {ACS Photonics},
   title = {Maximal Optical Chirality via Mode Coupling in Bilayer Metasurfaces},
   year = {2025},
}

@article{Guan2022,
   author = {Jun Guan and Jeong Eun Park and Shikai Deng and Max J.H. Tan and Jingtian Hu and Teri W. Odom},
   doi = {10.1021/acs.chemrev.2c00011},
      
   issue = {19},
   journal = {Chemical Reviews},
    
   pages = {15177-15203},
   pmid = {35762982},
   publisher = {American Chemical Society},
   title = {Light-Matter Interactions in Hybrid Material Metasurfaces},
   volume = {122},
   year = {2022},
}

@article{Review_USC,
  title={Ultrastrong coupling regimes of light-matter interaction},
  author={Forn-Díaz, P and Lamata, L and Rico, E and Kono, J and Solano, E},
  journal={Reviews of Modern Physics},
  volume={91},
  number={2},
  pages={025005},
  year={2019},
  publisher={APS}
}

@article{Babicheva2024,
   author = {Viktoriia E. Babicheva and Andrey B. Evlyukhin and Andrey B. Evlyukhin and Andrey B. Evlyukhin},
   doi = {10.1364/AOP.510826}, 
   issue = {3},
   journal = {Advances in Optics and Photonics},
   pages = {539-658},
    volume={16},
   publisher = {Optica Publishing Group},
   title = {Mie-resonant metaphotonics},
   year = {2024}
}

@article{Dovzhenko2018,
   author = {D. S. Dovzhenko and S. V. Ryabchuk and Yu P. Rakovich and I. R. Nabiev},
   doi = {10.1039/c7nr06917k},
      
   issue = {8},
   journal = {Nanoscale},
    
   pages = {3589-3605},
   pmid = {29419830},
   publisher = {Royal Society of Chemistry},
   title = {Light-matter interaction in the strong coupling regime: Configurations, conditions, and applications},
   volume = {10},
   year = {2018}
}

@article{Lepeshov2018,
   author = {Sergey Lepeshov and Yuri Kivshar},
   doi = {10.1021/acsphotonics.8b00246},
      
   issue = {7},
   journal = {ACS Photonics},
    
   pages = {2888-2894},
   publisher = {American Chemical Society},
   title = {Near-Field Coupling Effects in Mie-Resonant Photonic Structures and All-Dielectric Metasurfaces},
   volume = {5},
   year = {2018},
}

@article{Friedrich1985,
   author = {H Friedrich and D Wintgen},
   issue = {6},
   journal = {Physical Review A},
   title = {Interfering resonances and bound states in the continuum},
   volume = {32},
   year = {1985},
}

@article{Overvig2020,
  title={Selection rules for quasibound states in the continuum},
  author={Overvig, Adam C and Malek, Stephanie C and Carter, Michael J and Shrestha, Sajan and Yu, Nanfang},
  journal={Physical Review B},
  volume={102},
  number={3},
  pages={035434},
  year={2020},
  publisher={APS}
}

@article{CMT-OG,
   author = {Wonjoo Suh and Zheng Wang and Shanhui Fan},
   doi = {10.1109/JQE.2004.834773},
      
   issue = {10},
   journal = {IEEE Journal of Quantum Electronics},
    
   pages = {1511-1518},
   title = {Temporal coupled-mode theory and the presence of non-orthogonal modes in lossless multimode cavities},
   volume = {40},
   year = {2004}
}

@article{haus2002coupled,
  title={Coupled-mode theory},
  author={Haus, Hermann A and Huang, Weiping},
  journal={Proceedings of the IEEE},
  volume={79},
  number={10},
  pages={1505--1518},
  year={2002},
  publisher={IEEE}
}

@article{Mez-Espina2024,
   author = {Luis Manuel Máñez-Espina and Ihar Faniayeu and Viktar Asadchy and Ana Díaz-Rubio},
      
   issue = {1},
   journal = {Advanced Optical Materials},
    
   publisher = {John Wiley and Sons Inc},
   title = {Extreme Nonreciprocity in Metasurfaces Based on Bound States in the Continuum},
   volume = {12},
   year = {2024}
}

@article{vandeGroep2013,
   author = {J. van de Groep and A. Polman},
   doi = {10.1364/oe.21.026285},
      
   issue = {22},
   journal = {Optics Express},
    
   pages = {26285},
   publisher = {Optica Publishing Group},
   title = {Designing dielectric resonators on substrates: Combining magnetic and electric resonances},
   volume = {21},
   year = {2013}
}

@article{Alaee2015,
   author = {Rasoul Alaee and Mohammad Albooyeh and Aso Rahimzadegan and Mohammad S. Mirmoosa and Yuri S. Kivshar and Carsten Rockstuhl},
      
   issue = {24},
   journal = {Physical Review B},
    
   publisher = {American Physical Society},
   title = {All-dielectric reciprocal bianisotropic nanoparticles},
   volume = {92},
   year = {2015}
}

@article{Zhang2025,
   author = {Baohe Zhang and Anlong Dong and Junru Wang and Meng Qin and Jian Qiang Liu and Wen Huang and Shuyuan Xiao and Hongju Li},
   doi = {10.1021/acs.nanolett.5c00426},
      
   issue = {11},
   journal = {Nano Letters},
    
   pages = {4568-4575},
   pmid = {40066775},
   publisher = {American Chemical Society},
   title = {Strong Coupling and Electromagnetically Induced Transparency in Multiple-BIC-Driven Metasurfaces},
   volume = {25},
   year = {2025}
}

@article{Xie2021,
   author = {Peng Xie and Zhengchen Liang and Tongtong Jia and Daimin Li and Yixuan Chen and Peijie Chang and Hong Zhang and Wei Wang},
      
   issue = {12},
   journal = {Physical Review B},
    
   publisher = {American Physical Society},
   title = {Strong coupling between excitons in a two-dimensional atomic crystal and quasibound states in the continuum in a two-dimensional all-dielectric asymmetric metasurface},
   volume = {104},
   year = {2021}
}

@article{Zhen2014,
  title={Topological nature of optical bound states in the continuum},
  author={Zhen, Bo and Hsu, Chia Wei and Lu, Ling and Stone, A Douglas and Solja{\v{c}}i{\'c}, Marin},
  journal={Physical review letters},
  volume={113},
  number={25},
  pages={257401},
  year={2014},
  publisher={APS}
}

@Article{Hsu2016,
author={Hsu, Chia Wei
and Zhen, Bo
and Stone, A. Douglas
and Joannopoulos, John D.
and Solja{\v{c}}i{\'{c}}, Marin},
title={Bound states in the continuum},
journal={Nature Reviews Materials},
year={2016},
  
day={19},
volume={1},
number={9},
pages={16048},
doi={10.1038/natrevmats.2016.48},
}

@Article{Kang2023,
author={Kang, Meng
and Liu, Tao
and Chan, C. T.
and Xiao, Meng},
title={Applications of bound states in the continuum in photonics},
journal={Nature Reviews Physics},
year={2023},
day={01},
volume={5},
number={11},
pages={659-678},
doi={10.1038/s42254-023-00642-8},
}

@article{Marinica2008,
  title={Bound states in the continuum in photonics},
  author={Marinica, DC and Borisov, AG and Shabanov, SV},
  journal={Physical review letters},
  volume={100},
  number={18},
  pages={183902},
  year={2008},
  publisher={APS}
}

@article{Qin2023,
   author = {Meibao Qin and Junyi Duan and Shuyuan Xiao and Wenxing Liu and Tianbao Yu and Tongbiao Wang and Qinghua Liao},
      
   issue = {4},
   journal = {Physical Review B},
    
   publisher = {American Physical Society},
   title = {Strong coupling between excitons and quasibound states in the continuum in bulk transition metal dichalcogenides},
   volume = {107},
   year = {2023}
}

@article{Weber2023,
   author = {Thomas Weber and Lucca Kühner and Luca Sortino and Amine Ben Mhenni and Nathan P. Wilson and Julius Kühne and Jonathan J. Finley and Stefan A. Maier and Andreas Tittl},
   doi = {10.1038/s41563-023-01580-7},
      
   issue = {8},
   journal = {Nature Materials},
    
   pages = {970-976},
   pmid = {37349392},
   publisher = {Nature Research},
   title = {Intrinsic strong light-matter coupling with self-hybridized bound states in the continuum in van der Waals metasurfaces},
   volume = {22},
   year = {2023}
}

@article{Limonov2017,
   author = {Mikhail F. Limonov and Mikhail V. Rybin and Alexander N. Poddubny and Yuri S. Kivshar},
   doi = {10.1038/NPHOTON.2017.142},
      
   issue = {9},
   journal = {Nature Photonics},
    
   pages = {543-554},
   publisher = {Nature Publishing Group},
   title = {Fano resonances in photonics},
   volume = {11},
   year = {2017},
}

@article{Koshelev2018,
  title={Asymmetric metasurfaces with high-Q resonances governed by bound states in the continuum},
  author={Koshelev, Kirill and Lepeshov, Sergey and Liu, Mingkai and Bogdanov, Andrey and Kivshar, Yuri},
  journal={Physical review letters},
  volume={121},
  number={19},
  pages={193903},
  year={2018},
  publisher={APS}
}

\end{document}